\documentclass[prl,aps,floats,twocolumn,groupedaddress,preprintnumbers]{revtex4}

\usepackage[latin1]{inputenc}                    
\usepackage{graphicx}                            
\usepackage{latexsym}                            
\usepackage{amsfonts}                            
\usepackage{amssymb}                             
\usepackage{amsmath}                             
\usepackage[mathscr]{eucal}                      
\usepackage{dcolumn}                             
\usepackage{theorem}                             
\usepackage{footnote}
\usepackage{enumitem}
\usepackage{bm}
\usepackage{color}
\usepackage{placeins}

\usepackage{flushend}

\def\be{\begin{equation}}
\def\ee{\end{equation}}
\def\bea{\begin{eqnarray}}
\def\eea{\end{eqnarray}}
\def\bf{\begin{figure}}
\def\ef{\end{figure}}
\def\bc{\begin{center}}
\def\ec{\end{center}}

\def\fo{$F_1$ }

\def\q2{$Q^2$ }
\def\w2{$W^2$ }

\def\al{\alpha}
\def\bt{\beta}
\def\nn{\nonumber}

\begin{document}

\preprint{
\vbox{
\hbox{ADP-14-4/T862}
}}

\title{Momentum transfer dependence of the proton's electric and magnetic polarizabilities}
\author{N.~L.~Hall}

\author{A.~W.~Thomas}

\author{R.~D.~Young}

\affiliation{\mbox{ARC Centre of Excellence for Particle Physics 
	at the Terascale, and CSSM}, School of Chemistry and Physics, 
	University of Adelaide, Adelaide SA 5005, Australia}

\date{\today}

\begin{abstract}
The $Q^2$-dependence of the sum of the electric and magnetic polarizabilities of the proton is calculated over the range $0 \leq Q^2 \leq 6$ GeV$^2$ using the generalized Baldin sum rule. Employing a parametrization of the \fo structure function valid down to $Q^2 = 0.06$ GeV$^2$, the polarizabilities at the real photon point are found by extrapolating the results of finite $Q^2$ to $Q^2 = 0$ GeV$^2$. We determine the evolution over four-momentum transfer to be consistent with the Baldin sum rule using photoproduction data, obtaining $\alpha + \beta = 13.7 \pm 0.7 \times 10^{-4}\, \text{fm}^3$.
\end{abstract}

\pacs{14.20.Dh, 11.35.Hx, 13.60.Hb}

\keywords{Polarizability, Baldin sum rule}

\maketitle

\section{I. \; \, Introduction}

In order to study the internal structure of the nucleon, low-energy electron-proton scattering experiments are used to measure the response of the proton to an electromagnetic field. The electric ($\alpha$) and magnetic ($\beta$) polarizabilities are the fundamental parameters which describe this response. Physically, $\al$ and $\bt$ give a measure of the rigidity of the nucleon and are related to the unpolarized photoabsorption cross section by the Baldin sum rule \cite{Baldin1960310,Lap:13},
\be
	\alpha + \beta = \frac{1}{4 \pi^2} \int_{\nu_{\pi}}^{\infty} \frac{ \sigma_{1/2} + \sigma_{3/2}}{\nu^2} \;  d \nu,
	\label{eqn:BSR}	
\ee
where $\sigma_{\lambda}$ is cross section for the production of $\lambda = 1/2, 3/2$ helicity states, $\nu$ is the photon energy and $\nu_{\pi}$ the pion production threshold. 

Further insight into the spatial distribution of the polarizabilities in the proton is gained from an understanding of the evolution of $\al$ and $\bt$ as a function of the four-momentum transfer, $Q^2$. Extending the above relation to incorporate this $Q^2$-dependence leads to the generalized Baldin sum rule \cite{Drechsel2003}, 
\bea
	\alpha (Q^2) + \beta (Q^2) &=& \frac{8 \alpha_{\text{em}} M}{Q^4} \int_0^{x_\pi} \!x F_1 (x, Q^2) \,dx, \nn \\
							   &=& 8 \alpha_{\text{em}} M \int_{W_\pi^2}^{\infty} \frac{F_1 (W^2, Q^2)}{\left(W^2 - M^2 + Q^2\right)^3}\, dW^2 \! , \nn \\
	\label{eqn:gBSR}	
\eea
where $M$ is the mass of the proton, $\alpha_{\text{em}}$ is the fine structure constant, $x = Q^2/(W^2 - M^2 + Q^2)$ is the Bjorken scaling variable and $W$ the invariant mass. $F_1$ is the proton's electromagnetic structure function and we also note that in a similar manner to Eq. (\ref{eqn:BSR}), $x_\pi$ and $W_\pi$ refer to the pion production point. 

There has been considerable interest of late in accurately determining, both experimentally  \cite{Liang:2004tk} and theoretically \cite{Sibirtsev2013, Lensky:2014efa}, the $Q^2$-dependence of $\al$ and $\bt$. In Ref.~\cite{Liang:2004tk}, Liang {\it et al.} exploited $e-p$ scattering data from the JLab E94-110 experiment \cite{Liang:2004tj} to obtain $\alpha (Q^2) + \beta (Q^2)$ in the range $0.3 < Q^2 < 4$ GeV$^2$. More recently, Sibirtsev and Blunden (SB) made use of updated data \cite{Liang:2013tj} from the same experiment to construct a parametrization of the electromagnetic \fo structure function \cite{Sibirtsev2013,Sibirtsev2010} which they then used to evaluate the Baldin integral. By doing so, they were able to provide valuable information on the low $Q^2$ properties of the electric and magnetic polarizabilities. Additionally, they found that the resonance region contribution dominates for $Q^2 < 1$ GeV$^2$ \cite{Sibirtsev2013}.

In this report we improve upon the preceding work, utilizing the Adelaide-Jefferson Lab-Manitoba (AJM) parametrization of \fo which is consistent down to much lower momentum transfer $\sim 0.06$ GeV$^2$. After a brief description of this 
construction in Sec.~II, we evaluate the generalized Baldin sum rule, presenting our results in Sec.~III. A discussion of their relation to those of Ref.~\cite{Sibirtsev2013} is also included in this section, whilst final conclusions are given in Sec.~IV.

%
\section{II. \; \,Adelaide-Jefferson Lab-Manitoba parametrization}
\label{sec:AJM}

The parametrization of the \fo structure function we employ in Eq.~(\ref{eqn:gBSR}) was previously developed in Ref.~\cite{Hall2013} for parity-violating asymmetry calculations. In this earlier work, the $Q^2$-$W^2$ plane of the structure function was separated into distinct regions according to the physics most appropriate to that region. At the low $Q^2$ and $W^2$, designated `Region I', \fo was given by Christy and Bosted's (CB) fit \cite{Christy2010} to data from a 2008 version of Ref.~\cite{Liang:2004tj}, with uncertainties of 3-5\%. For the low $Q^2$ but high $W^2$ range (`Region II'), a Regge parametrization was used in combination with the vector meson dominance model \cite{Sakurai1972a,Alwall2004}. Finally, the high $Q^2$ and $W^2$ region was described by parton distribution functions given by Alekhin {\it et al.} \cite{Alekhin2012a}. 

One additional constraint of the AJM parametrization was that at the boundary between the regions, the structure functions were required to match onto each other. Since the individual descriptions have areas in the $Q^2$-$W^2$ plane where they overlap, the position of the `hard' borders should have little effect on the end result. 

\begin{figure*}[t]
\begin{center}
\includegraphics[width=0.81\textwidth]{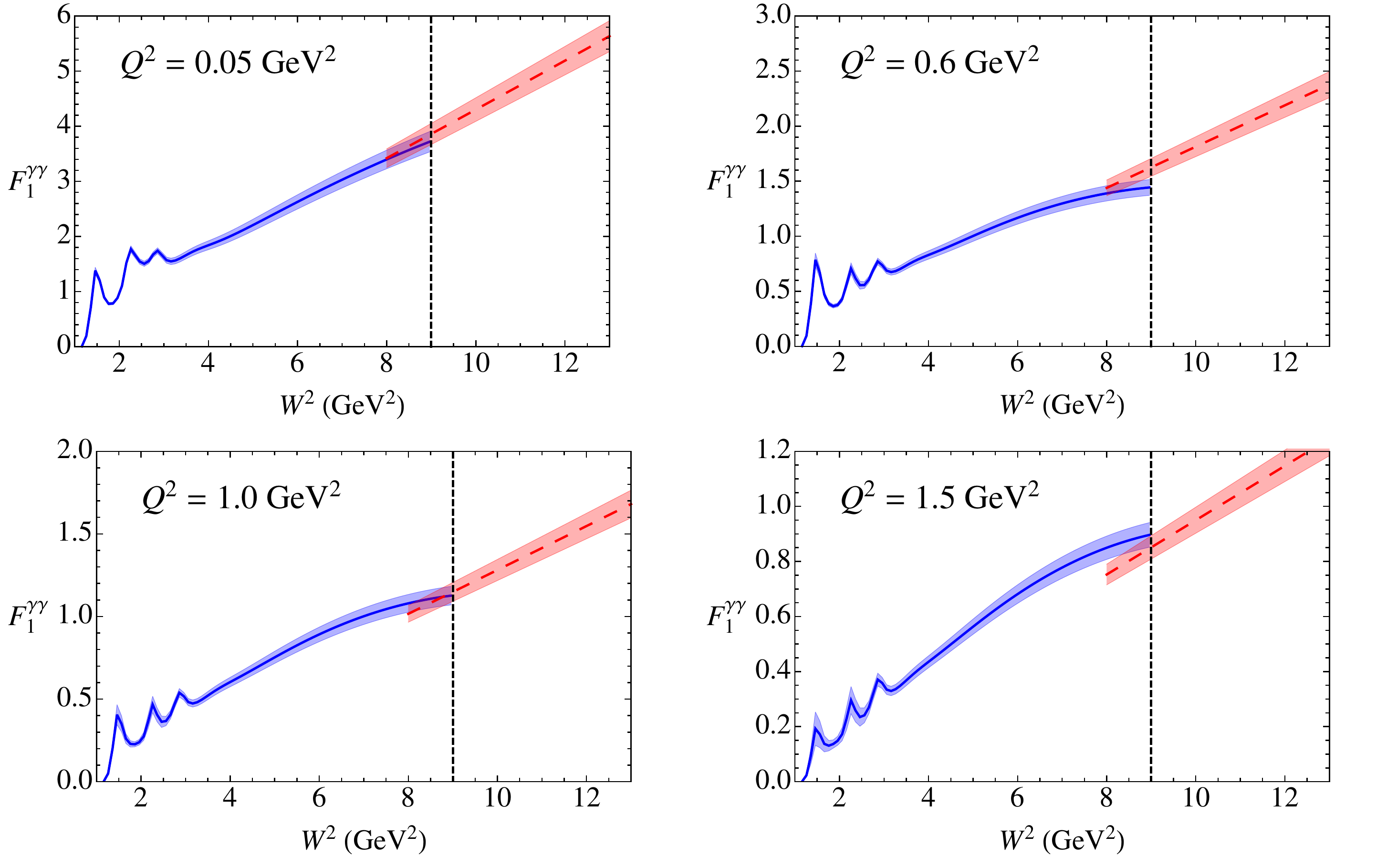}
\caption{(color online)
	Proton \fo structure function versus $W^2$ at fixed
	$Q^2=0.05$, 0.6, 1.0 and 1.5~GeV$^2$ for the CB fit
	\cite{Christy2010} at low $W^2$ (blue solid) and VMD+Regge
	parametrization \cite{Alwall2004} at high $W^2$ (red dashed).
	The boundary between these is indicated by the vertical dashed
	line at $W^2 = 9$~GeV$^2$.}
\label{fig:F1CV}
\end{center}
\end{figure*}

In Fig.~\ref{fig:F1CV} we plot $F_1 (W^2, Q^2) $ as a function of $W^2$ for multiple values of $Q^2$ ranging from $Q^2 = 0.05$ to $1.5$ GeV$^2$. It is clear from these plots that the two descriptions of the structure function are in good agreement at the boundaries. The errors, shown by the shaded bands, come from assigning a conservative 5\% uncertainty to both parametrizations of the structure function.

%
\section{III. \; \,$Q^2$ dependence of $\al$ and $\bt$}
\label{sec:Pols}

As we would like to determine the evolution of the polarizabilities all the way to the photoproduction point, the behavior of the parametrization of \fo at very low $Q^2$ is of particular importance. CB's fit includes data to $Q^2$ as low as $0.06$ GeV$^2$ as well as photoproduction data, showing good agreement with the experimental values in the regions where there is data. However, as $Q^2 \rightarrow 0$  the parametrization fails to accurately match onto the real photon point as a result of difficulties in correctly fitting the second resonance peak. In order to get around this problem, we determine $\al(Q^2) + \bt(Q^2)$ down to $Q^2 = 0.06$ GeV$^2$ using CB's $F_1$, before extrapolating the results the rest of the way to $Q^2 = 0$ GeV$^2$. By fitting the results over several ranges of momentum transfer, we qualify any systematic uncertainty in this extrapolation.

\begin{figure}
\begin{center}
\includegraphics[width=\columnwidth]{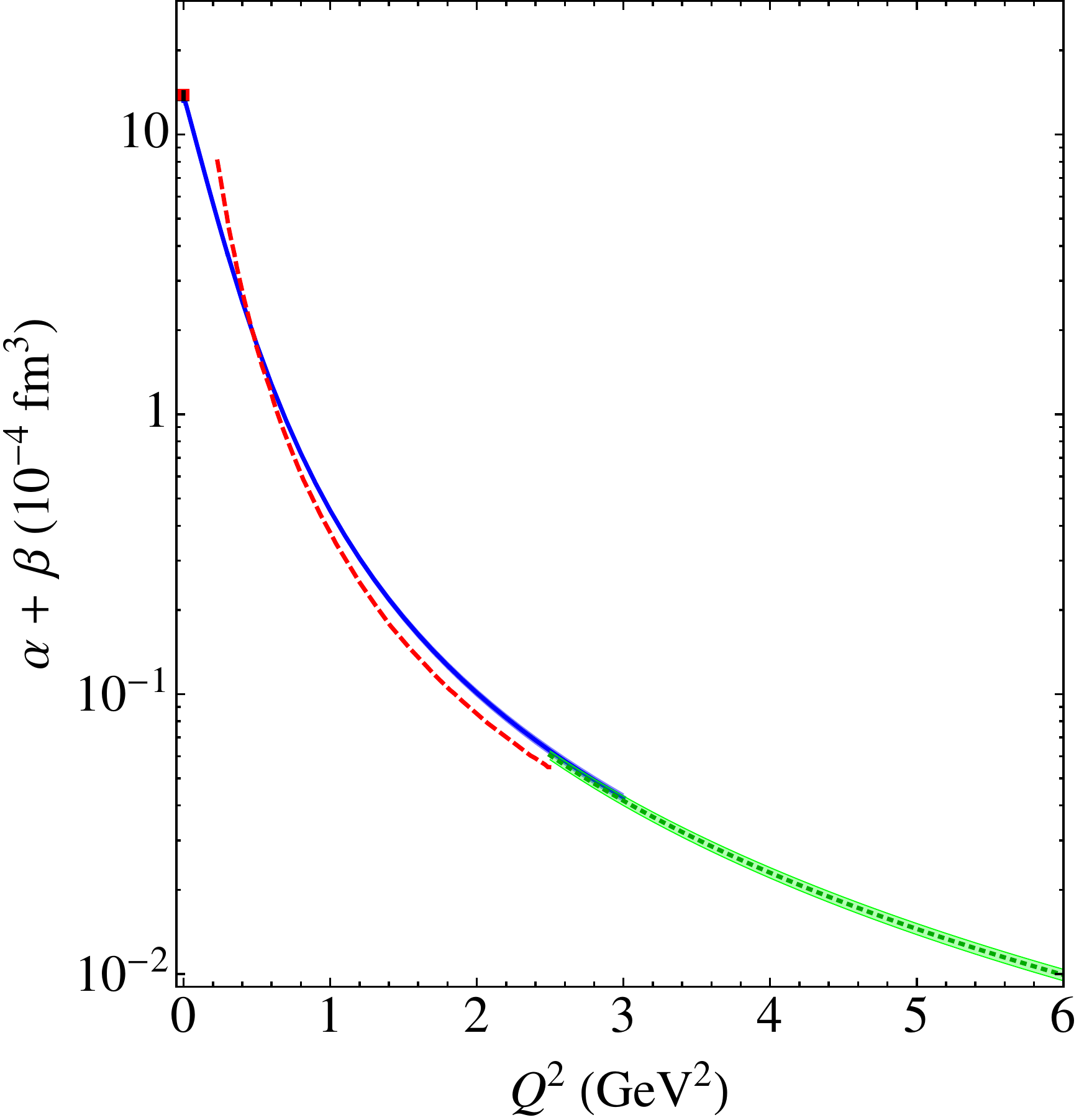}
\caption{(color online) The sum of the electric and magnetic polarizabilities as a function of $Q^2$. The blue line (with errors) represents our calculation for $Q^2 \leq 2.5$ GeV$^2$, the green dotted line for $Q^2 > 2.5$ GeV$^2$, whilst the red-dashed line is the earlier calculation of Ref.~\cite{Sibirtsev2013}. On this scale the values of $\al + \bt$ at the real photon point overlay each other and are given by the black triangle \cite{Babusci1998} and red square \cite{OlmosdeLeon:2001zn}.
}
\label{fig:albe}
\end{center}
\end{figure}

In Fig.~\ref{fig:albe} we evaluate the generalized Baldin sum rule over the range $ 0.06 \leq Q^2 \leq 6$ GeV$^2$, also including the results of Ref.~\cite{Sibirtsev2013}. Although we show the $Q^2$-dependence up to $6$ GeV$^2$, the determination of the polarizabilities is limited only by the accuracy and validity of the PDFs. In principal, one could calculate the sum all the way up to LHC energies. 

\begin{figure}
\begin{center}
\includegraphics[width=0.80\columnwidth]{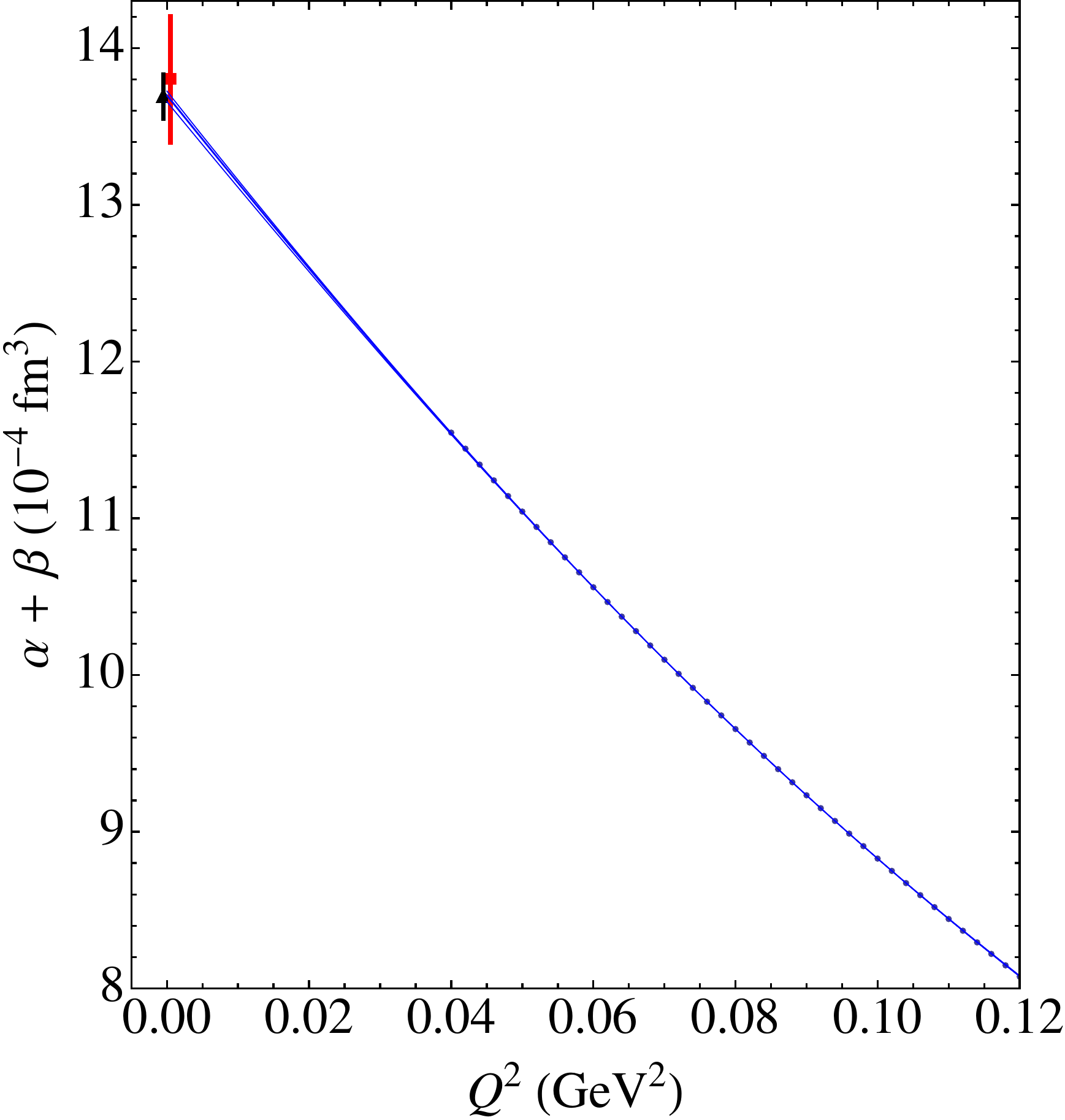}
\caption{(color online) In this plot, the individual points are values extracted using the AJM parametrization, whilst the blue lines are fits to points over four distinct ranges in $Q^2$. The values of $\al + \bt$ at $Q^2 = 0$ GeV$^2$ have been offset slightly and are given by the black triangle \cite{Babusci1998} and red square \cite{OlmosdeLeon:2001zn}.
}
\label{fig:ZE}
\end{center}
\end{figure}

For the low-$Q^2$ behavior, as mentioned before, we fit the results for the Baldin integral (in this case using the inverse of a polynomial) over four different ranges of $Q^2$:
(I) $0.04<Q^2<0.10$ GeV$^2$; 
(II) $0.04<Q^2<0.12$ GeV$^2$; 
(III) $0.06<Q^2<0.10$ GeV$^2$;
(IV) $0.06<Q^2<0.12$ GeV$^2$.
These fits were then extrapolated to $Q^2 = 0$ GeV$^2$, with the results of the extrapolations in Fig.~\ref{fig:ZE} showing minimal variation of $\al + \bt$ over the $Q^2$ ranges.

At the real photon point we obtain, 
\be
 \alpha + \beta = (13.7 \pm 0.7) \times 10^{-4}\; \text{fm}^3
\ee
where we quote results for the fit over the region (III). The error $\pm 0.7$ comes from the conservative 5\% uncertainty associated with the CB parametrization. Our result is in excellent agreement with the previous determination of Babusci {\it et al.} \cite{Babusci1998}, who quote
\be
 \alpha + \beta = (13.69 \pm 0.14) \times 10^{-4}\; \text{fm}^3
\ee
and the more recent value,
\be
  \alpha + \beta = (13.8 \pm 0.4) \times 10^{-4}\; \text{fm}^3,
\ee
of Ref. \cite{OlmosdeLeon:2001zn}.
 
As can be seen, our evaluation differs substantially from that of Ref.~\cite{Sibirtsev2013}. Most notably, as $Q^2$ tends to $0$ the sum of $\al(Q^2)$ and $\bt(Q^2)$ clearly converges to the value at the real photon point, whereas an extrapolation of the SB curve would overshoot this value significantly. The variation between the two results may be explained by the fact that the AJM parametrization uses the fit of CB for \fo which is consistent to much lower $Q^2$ values. Although not shown on the graph and in agreement with SB, we found that the resonance contribution dominates at low $Q^2$. (We also point out that varying the $W^2$ boundary between Region I and II has a negligible effect on the final values.)

Additionally, we calculate the ``radius" of the sum of the polarizabilities, i.e.,
\be
 \langle r^2 \rangle = \left. \frac{-6}{H(0)} \frac{d H(Q^2)}{d Q^2} \right|_{Q^2=0}
\ee
where in this case, $H(Q^2) = \al(Q^2) + \bt(Q^2)$. For the fit used to determine the small-$Q^2$ extrapolation the radius is 
\be
 \langle r^2 \rangle_{\al + \bt}^{1/2} = 0.98 \pm 0.05 \, \text{fm}.
\ee
%

%
%

%

\section{IV. \; \,Conclusion}
\label{sec:Con} 
Utilizing a description of the \fo structure function which is consistent down to $0.06$ GeV$^2$ and extrapolating to $Q^2 = 0$ GeV$^2$, we have shown that the $Q^2$-dependence of the electric and magnetic polarizabilities matches accurately onto the dispersion relation at the real photon
\hspace{-4.6mm} point. We explain the difference from the results of SB at low $Q^2$ from this property of our structure function. At higher $Q^2$, the difference arises primarily from the deviation of their model from the CB parametrization. Increased data at very low $Q^2$ and low $W^2$ region would be useful in decreasing the uncertainties still further. 

\section*{Acknowledgements}
We are pleased to thank P. Blunden and W. Melnitchouk for their role in the development of the AJM model. 
This work was supported by the University of Adelaide and the Australian Research Council through the ARC Centre of Excellence for Particle Physics at the Terascale and grants FL0992247 (AWT), DP140103067, FT120100821 (RDY).


%
%
\bibliographystyle{apsrev}
\bibliography{Polarbib}

\end{document}